\documentclass[12pt,headings=big,numbers=noenddot,DIV=14,a4paper]{article}%

\pdfoutput=1

\usepackage[margin=1 in]{geometry}
\usepackage{comment}
\usepackage{graphicx}  
\usepackage{dcolumn}   
\usepackage{bm,relsize}        
\usepackage{amsfonts,amsmath,amssymb,amsthm,mathtools}
\usepackage{slashed}
\usepackage{placeins}
\usepackage{lipsum, color}
\usepackage[usenames,dvipsnames,svgnames,table]{xcolor}
\usepackage[linktoc=page,bookmarks=false,colorlinks=true,linkbordercolor=RoyalBlue,citebordercolor=ForestGreen,urlbordercolor=CornflowerBlue]{hyperref}
\hypersetup{
    colorlinks=true,
    linkcolor=ForestGreen,
    filecolor=ForestGreen,      
    urlcolor=ForestGreen,
    citecolor=ForestGreen,
}
\usepackage[sort&compress,numbers,merge]{natbib}
\usepackage{afterpage}
\usepackage{floatpag}
\usepackage{tikz-feynman}
\usepackage{mathtools}

\def\be{\begin{equation}}
\def\ee{\end{equation}}
\newcommand{\ba}{\begin{array}}
\newcommand{\ea}{\end{array}}

\usepackage[normalem]{ulem}

\definecolor{mtcolor}{rgb}{.8,.3,.1}

\begin{document}

\color{black}

\begin{center}

{\LARGE \bf
Searching for inelastic dark matter \\with future LHC experiments
\large
}

\medskip
\bigskip\color{black}\vspace{0.5cm}

{
{\large Enrico Bertuzzo},$^a$
{\large Andre Scaffidi},$^{b}$
{\large Marco Taoso}$^{b}$\,\footnote{E-mail: \href{mailto:bertuzzo@if.usp.br}{bertuzzo@if.usp.br}, \href{mailto:andre-joshua.scaffidi@to.infn.it}{andre-joshua.scaffidi@to.infn.it}, \href{mailto:marco.taoso@to.infn.it}{marco.taoso@to.infn.it}}
}
\\[7mm]

{\it \small $^a$ Instituto de F\'{i}sica, Universidade de S\~{a}o Paulo, C.P. 66.318, 05315-970 S\~{a}o Paulo, Brazil}\\
{\it \small $^b$ I.N.F.N. sezione di Torino, via P. Giuria 1, I-10125 Torino, Italy}\\
\end{center}

\bigskip

\centerline{\bf Abstract}
\begin{quote}
\color{black}

We consider a dark sector containing a pair of almost degenerate states coupled to the Standard Model through a dark photon mediator. This set-up constitutes a simple realization of the inelastic dark matter scenario. The heaviest dark state is long-lived, in the limit of a small kinetic mixing among the dark photon and the Standard Model hypercharge gauge boson, and/or of a small mass splitting among the dark states. 
We study the prospects for detection of this scenario at proposed LHC experiments dedicated to search for long-lived particles, namely FASER, MATHUSLA, CODEX-b, AL3X, MAPP, ANUBIS and FACET.
We consider both the cases of fermionic and scalar inelastic dark matter.
We show that these experimental facilities can probe unexplored regions of the parameter space of this model, and we highlight their complementary roles.

\medskip

\end{quote}

\clearpage
\noindent\makebox[\linewidth]{\rule{\textwidth}{1pt}} 
\tableofcontents
\noindent\makebox[\linewidth]{\rule{\textwidth}{1pt}} 	
\clearpage

\section{Introduction}
\label{sec:intro}

In 
recent years, a rich experimental program has been put forward to search for Long Lived Particles (LLPs)~\cite{Alimena:2019zri,Lee:2018pag}. In accelerator experiments LLPs travel macroscopic distances before decaying, leading to signatures such as displaced vertices or missing energy. To fully exploit the potential of the Large Hadron Collider (LHC), new experimental facilities dedicated to search for LLPs have been recently proposed, namely FASER\,\cite{Feng:2017uoz,FASER:2019aik}, MATHUSLA\,\cite{Chou:2016lxi,Curtin:2018mvb,MATHUSLA:2020uve}, CODEX-b\,\cite{Gligorov:2017nwh,Aielli:2019ivi}, AL3X,\cite{Gligorov:2018vkc}, MAPP\,\cite{Staelens:2019gzt,Pinfold:2019zwp}, ANUBIS\,\cite{Bauer:2019vqk} and FACET\,\cite{Cerci:2021nlb}.
These proposed detectors would be located near the LHC Interaction Points (IP), and
are designed to detect the decays of LLPs produced at the LHC, having at the same time the capability to dramatically reduce backgrounds.
Interestingly, these projects are highly complementary to the existing LHC experiments ATLAS, CMS, and LHCb.  

LLPs are predicted in many extensions of the Standard Model (SM) of particle physics, see e.g.\,\cite{Curtin:2018mvb,Alexander:2016aln,Beacham:2019nyx}. In particular, they often arise in dark sectors, which are scenarios postulating the existence of new states, possibly including Dark Matter candidates (DM), feebly coupled to the SM. In this work, we focus on the possibility of new SM singlets in the form of inelastic DM (iDM)\,\cite{Tucker-Smith:2001myb}.
This model contains two particles almost degenerate in mass, the lightest one being the DM candidate (of course in general an extension with more than two states is possible). If this pair of particles is then coupled to the SM only through tiny portal interactions, the decay of the heaviest dark state is suppressed and this state acts as a LLP. iDM is particularly attractive from the phenomenological point of view, because it allows to explain the cosmological DM abundance via the standard thermal freeze-out and, at the same time, evade existing bounds.
Coannihilations of the iDM particles in the early Universe determine their relic abundance. On the other hand, these processes are suppressed at later times, allowing to fulfill the constraints from indirect DM searches and Cosmic Microwave Background (CMB) observations, which are particularly tight for DM masses $\lesssim {\rm few}\,\mathcal{O}(10)$ GeV. The inelastic scattering of DM into nuclei is also suppressed, for large enough mass splitting among the dark states. Therefore, also direct detection constraints can be accommodated in this scenario.
In general, the portal interactions among the dark sector and the SM can be described in terms of renormalizable (see \,\cite{Curtin:2018mvb,Alexander:2016aln,Beacham:2019nyx} and references therein) or non-renormalizable portal operators\,\cite{Darme:2020ral,Bertuzzo:2020rzo,Contino:2020tix}.
We adopt the first option, considering a mediator between the SM and the dark sector in form of a dark photon\,\cite{Berlin:2018jbm,Izaguirre:2015zva}. A relatively light mediator is required to explain the DM abundance via standard thermal freeze-out processes. Building upon ref.\,\cite{Berlin:2018jbm}, we perform an analysis of this iDM model, focusing on future searches of LLPs at the LHC.
Our goal is to extend the analysis of\,\cite{Berlin:2018jbm} to recently proposed experimental facilities~\footnote{We also update the geometry of the FASER and MATHUSLA detectors to the ones most recently considered by the experimental collaborations.}.
In particular, we determine the region of parameter space that can be explored by these proposed future LHC detectors, highlighting their complementary role in covering the parameter space of iDM models.

The paper is organized as follows. In Section\,\ref{sec:model} we introduce the iDM models under consideration. We will consider both scalar and fermionic iDM states. In Section\,\ref{sec:setup} we describe how the projected sensitivities are computed. We present our main results in Section\,\ref{sec:results} and we offer our conclusions in Section\,\ref{sec:conclusions}.

\section{Inelastic dark matter}
\label{sec:model}
Models of iDM have been proposed long ago as a simple way to evade the strong constraints coming from direct detection experiments~\cite{Han:1997wn,Hall:1997ah}, and have later been used to reconcile the observations by the DAMA experiment with the exclusion limits derived from other direct detection experiments~\cite{Tucker-Smith:2001myb}~\footnote{The mass splitting considered for this purpose is typically of $\mathcal{O}(100)\, {\rm keV}$, much smaller than the ones considered here.}. Although recent analysis seems to exclude an iDM interpretation of the DAMA signal~\cite{Jacobsen:2021vbr}, in recent years
iDM has been studied as a paradigmatic example of a dark sector that can evade the stringent bounds coming from direct detection~\cite{XENON:2018voc} and CMB observations~\cite{Planck:2018vyg}, with DM particles still being thermally produced.

To illustrate the main idea, let us consider a scenario where, at leading order, the interaction between the dark and SM sectors occours via a trilinear vertex of the form $D_1$-$D_2$-mediator, where $D_2$ is a dark state obeying $m_{2} \gtrsim m_{1}$ (with $m_{1}$ and $m_{2}$ the $D_1$ and $D_2$ masses, respectively) and $D_1$ is the DM candidate. 
The mediator involved in this vertex is coupled to SM particles, providing therefore a portal between the dark sector and the SM.
In an underground laboratory experiment,
for a sufficiently large mass splitting 
\be
\Delta = \frac{m_2 - m_1}{m_1}, 
\ee
the incoming DM particle $D_1$ does not have enough energy to upscatter 
off a nucleus
into the heavier state $D_2.$
Therefore bounds from direct detection experiments are evaded. 
At the same time, the iDM model can also fulfill the constraints on DM annihilations derived from CMB data.
In fact, in the early Universe,
the cross section for co-annihilation
of iDM pairs $D_1\,D_2$
into 
SM particles is suppressed by an exponential term $\exp[-(m_2-m_1)/T]$~\cite{Griest:1990kh}.
These co-annihilation processes are therefore irrelevant at late times, after the recombination era,
as long as $(m_2-m_1)/T_{rec} \gg 1$, where $T_{rec}$ is the temperature at which recombination happens. For weak-size interactions, any mass splitting $\gg \mathcal{O}(10)$ eV
is sufficient to 
avoid the CMB bounds. More generically, for sufficiently large mass splitting, the constraints from indirect detection of DM are fulfilled in iDM models. 
On the other hand, as long as the mass splitting is not 
too large, 
the exponential factor 
is of ${\cal O}(1)$ at temperatures $T \sim m_1$, allowing for DM thermal production in the early Universe. 

Relevant constraints might arise from annihilations of $D_1$ pairs, for instance processes such as $D_1 D_1 \leftrightarrow f\bar{f},$ where $f$ are SM fermions. Similarly, also interactions mediating the $D_1$ elastic scattering off nuclei might be present.
These contributions are model dependent and will be discussed below. 

As we will see in a moment, the situation presented above can be realized in a simple way for both fermionic and scalar dark states. For definiteness, we will consider the case in which the mediator between the dark sector and the SM is the so-called dark photon~\cite{Holdom:1985ag}, i.e. the (massive) gauge boson associated with an additional $U(1)'$ symmetry, that interacts with the SM via a kinetic mixing term:
\be
{\cal L}_{int} = \frac{\epsilon}{2 \cos\theta_w} A_{\mu\nu}' B^{\mu\nu} .
\ee
In the previous equation, $B_{\mu\nu}$ and $A'_{\mu\nu}$ are the field strengths of the gauge bosons associated with the hypercharge $U(1)_Y$ and the additional $U(1)'$, respectively, $\cos\theta_w$ is the cosine of the weak angle and $\epsilon$ is the kinetic mixing parameter. In the remainder of the paper, we will remain agnostic about the origin of the dark photon mass, supposing that any scalar degree of freedom associated with the spontaneous breaking of $U(1)'$ is sufficiently decoupled from the $D_1$, $D_2$ and $A'$ system. In the limit $\epsilon \ll 1$ and $m_{A'} \ll m_Z$ (with $m_{A'}$ and $m_Z$ the dark photon and $Z$ boson masses, respectively), all SM fermions acquire a coupling to the dark photon of the form~\cite{Curtin:2014cca}
\be
\label{eq:DPcoupling}
{\cal L}_{int} = e\, \epsilon A_\mu' \sum_f \bar{f} Q_f \gamma^\mu f, 
\ee
where $e\,Q_f$ is the electric charge of the SM fermion $f$.
This corresponds to the limit of photon-like interactions of the dark photon. 
However, since in our analysis we scan over the dark photon mass, considering also cases with a significant mixing between the $A'$ and $Z$ bosons, we use the full expressions for the couplings of $A'$ to the SM states and of $Z$ to the dark states\,\cite{Curtin:2014cca}.

\subsection{Fermionic case}
\label{sec:FIDM_summary}
We now turn to a brief discussion of two realizations of the inelastic dark matter scenario described above. We start with the case of fermionic dark states. Let us take two Weyl fermions $\psi_1$ and $\psi_2$ with charges $+1$ and $-1$ under $U(1)'$, respectively. The relevant Lagrangian besides standard kinetic terms (and written using two-component spinors) is given by
\be
{\cal L} =  - g_d \, A_\mu' \left( \psi_1^\dag \bar{\sigma}^\mu \psi_1 - \psi_2^\dag \bar{\sigma}^\mu \psi_2 \right) - \left(  m_D \psi_1 \psi_2 + \frac{\delta m_1}{2} \psi_1 \psi_1 + \frac{ \delta m_2}{2} \psi_2 \psi_2 + h.c \right),
\ee
where $g_d$ is the gauge coupling corresponding to $U(1)'$ and the Majorana masses $\delta m_{1,2}$ explicitly break $U(1)'$. The latter can be generated, for instance, by the vacuum expectation value of some scalar state (in the minimal model, the same one that generates the dark photon mass). Notice that, since in the $\delta m_{1,2} \to 0$ limit the $U(1)'$ symmetry is recovered in the fermionic sector, it is technically natural to have $\delta m_{1,2} \ll m_D$. In this limit the heavier state will typically be long lived. 
Taking all the mass parameters real for simplicity,
the mass eigenvalues at leading order in $\left(\frac{\delta m_1-\delta m_2}{m_D}\right)$
are 
\be
m_{1,2} = m_D \mp \frac{\delta m_1 + \delta m_2}{2}, 
\ee 
corresponding to the mass eigenstates 
\be
\chi_- =   i\frac{\psi_2 - \psi_1}{\sqrt{2}}+ i\frac{\delta m_1 - \delta m_2}{4 m_D} \frac{\psi_1 + \psi_2}{\sqrt{2}} , ~~~~~~ 
\chi_+ = \frac{\psi_1 + \psi_2}{\sqrt{2}}+\frac{\delta m_1 - \delta m_2}{4 m_D} \frac{\psi_1 - \psi_2}{\sqrt{2}} . 
\ee
The factor of $i$ in the definition of $\chi_-$ amounts to a choice of phase that automatically guarantees $m_1 > 0$ under the assumption $m_D >0$. With this choice of phase, the interactions with the dark photon can be expressed in terms of Majorana spinors $\chi_1 = (\chi_-, \chi_-^\dag)$ and $\chi_2 = (\chi_+, \chi_+^\dag)$, obtaining
\be\label{eq:Lint_f}
{\cal L}_{int}^\chi = i g_d \, \bar{\chi}_2 \gamma^\mu \chi_1 A_\mu' + {\cal O}\left(\frac{\delta m_1-\delta m_2}{m_D}\right) .
\ee
This is precisely a vertex of the form $D_1$-$D_2$-mediator described above, with $D_1 = \chi_1$ and $D_2 = \chi_2$. The terms suppressed by $(\delta m_1-\delta m_2)/m_D$ contain diagonal couplings of the type $\bar{\chi}_i \gamma^\mu \gamma_5\chi_i A_\mu '$, with $i=1,2$. 
In the limit of photon-like couplings as in eq.\,(\ref{eq:DPcoupling}), the annihilation cross-section for the process $\chi_1 \chi_1 \leftrightarrow f \bar{f}$ induced by these terms is of p-wave type, and thus it is typically sufficiently suppressed to fulfil the CMB bounds mentioned above.
In the following, we will assume that these terms are negligible or simply not present (implicitly taking 
the limit $(\delta m_1 -\delta m_2) \rightarrow 0 $).  
Notice that loop processes can induce $\chi_1$ elastic scattering off nucleons, as well as $\chi_1$ pair annihilations.
These processes are however suppressed, below the sensitivity of current probes\,\cite{Berlin:2018jbm}.

\subsection{Scalar case}
\label{sec:SIDM_summary}
A similar reasoning can be applied to the case in which the dark sector is populated by scalar states. Considering a complex scalar $\phi = (\phi_2 + i \phi_1)/\sqrt{2}$ with charge $+1$ under $U(1)'$, the relevant Lagrangian contains the terms: 
\be
{\cal L} = ig_d\, \left( \partial^\mu \phi^\dag \phi - \phi^\dag \partial^\mu \phi \right) A_\mu ' - m^2 \phi^\dag \phi - \left(\frac{\delta m^2}{2}\,\phi^2 + h.c. \right)
\ee
where the mass term $\delta m^2$ explicitly breaks $U(1)'$ and splits the real scalars $\phi_1$ and $\phi_2$, which acquire squared masses $m^2 \mp \delta m^2$, respectively. As in the previous case, it is technically natural to have $\delta m^2 \ll m^2$ and for $\phi_2$ to be long lived. As for the interactions with the dark photon, in terms of the mass eigenstates $\phi_{1,2}$ they become
\be\label{eq:Lint_sc}
{\cal L}_{int}^\phi = g_d \left( \partial^\mu \phi_1 \phi_2 - \phi_1 \partial^\mu \phi_2 \right) A_\mu',
\ee
once more of the form $D_1$-$D_2$-mediator with $D_1 = \phi_1$ and $D_2 = \phi_2$. In this case, the annihilation cross-section $\phi_1 \phi_1 \leftrightarrow f \bar{f}$, as well as the $\phi_1$ elastic scattering off nuclei,  are generated at the 1-loop level and will thus be parametrically suppressed by a factor $(\epsilon/(4\pi))^4$, sufficient to avoid the present constraints.

\medskip

In what follows we will use the interaction Lagrangians of Eqs.~\eqref{eq:Lint_f} and~\eqref{eq:Lint_sc} for our phenomenological analysis. 
In the discussion above and in the following, we consider the case of a dark photon heavier than the iDM pair, i.e. $m_A^{\prime}>m_1+m_2.$ The opposite limit corresponds to the secluded dark matter scenario\,\cite{Pospelov:2007mp}, which has a different phenomenology. In that case, at colliders, which is the main focus of our work, $A^{\prime}$ will decay into SM particles, rather than iDM pairs. Moreover, in the early Universe, the DM abundance is controlled by the annihilations $D_1\,D_1\rightarrow A^{\prime }A^{\prime }.$ These processes are strongly constrained by CMB and indirect detection bounds, see e.g.\,\cite{Cirelli:2016rnw,Bringmann:2016din}.

Before concluding this section, we discuss current experimental bounds on the dark photon parameter space. Depending on the relative importance of $\epsilon$ and $g_d$, the dark photon may decay mainly visibly (into SM particles) or invisibly (into dark sector particles)~\cite{Fabbrichesi:2020wbt}. Since we are interested in the case in which the interactions between $A'$ and the dark pair $D_1 D_2$ are sufficiently strong to allow for DM thermal production, in this work we consider only the so-called ``invisible'' dark photon. Focussing on the region $m_{A'} \gtrsim 1$ MeV, the most stringent bounds come from the dedicated search at BaBar~\cite{BaBar:2017tiz} and electroweak precision measurements at LEP~\cite{Curtin:2014cca}. The BaBar limits dominate in the region $m_{A'} \lesssim 8$ GeV, constraining $\epsilon \lesssim 10^{-3}$ in most of the parameter space and becoming as strong as $\epsilon \lesssim 4\times 10^{-4}$ for masses $m_{A'} \sim (4\div 6)$ GeV. On the other hand, LEP bounds dominate for larger $A^{\prime}$ masses and, as long as $m_{A'} \neq m_{Z}$, put a milder bound of $\epsilon \lesssim 2\times 10^{-2}$. Around $m_{A'} \simeq m_Z$ the limit becomes stronger, improving by roughly one order of magnitude. In addition, in this region the mixing between the $A'$ and $Z$ bosons is close to maximal, with the $Z$ boson potentially decaying into the dark states. For this reason, we also consider the bounds from $Z \to$ invisible\,\cite{Zyla:2020zbs}.

\section{Experimental setup and sensitivity estimation}
\label{sec:setup}

Our analysis focuses on a scenario with (fermionic or scalar) iDM and with a dark photon mediator heavier than the pair of dark states, i.e. $m_{A'}>m_1+m_2.$
As explained in the previous section, this construction allows the evasion of the stringent bounds from direct detection, indirect DM searches, and CMB observations. Furthermore, we are going to consider masses of the dark sector particles heavier than few GeVs. As we will see later, this region is favored by considerations on the DM relic abundance.
The dominant production mechanism of iDM at LHC are Drell-Yan processes with a $A'$ or $Z$ boson in the s-channel. An additional channel for the production of $A'$ particles, which would then decay into iDM pairs, is bremsstrahlung. This is relevant for $m_{A'}\lesssim$ few GeVs, and it is therefore subdominant in the range of masses than we explore.  
We implement the iDM models in \verb!Feynrules!\,\cite{Degrande:2014vpa} modifying the \verb!Feynrules!\ model file of\,\cite{Curtin:2014cca}, and use \noindent\verb!MadGraph5_aMC@NLO!\,\cite{Alwall:2014hca} to simulate $p p \to D_1 D_2$ events at $\sqrt{s} = 14$ TeV (where $D_i=\chi_i$ or $D_i=\phi_i$, $i=1,2$).

\subsection{Forecasting sensitivity}
The number of signal events in the detectors introduced in Sec.\,\ref{sec:intro}, can then be computed by multiplying the total number of $D_2$ particles produced at the LHC, by the probability that the $D_2$ particles decay inside the detector and lead to signal events, $f_{\rm signal}$.
The first term can be simply obtained from the production cross-section of $D_1 D_2$ pairs, that we evaluate with \verb!MadGraph5_aMC@NLO!, and the total integrated luminosity received by a given experiment, 
listed in Sec.\,\ref{sec:experiments}.
Instead, the quantity $f_{\rm signal}$ can be computed as:
\be\label{eq:Nsignal}
f_{\rm signal} = \left\langle \,\, f_{\rm dec}\, \epsilon_{\rm det} \,\, \right\rangle\ ,
\ee
where $f_{\rm dec}$ corresponds to the probability for a $D_2$ particle to decay inside the detector and it is given by:
\be\label{eq:Ndec}
f_{\rm dec} =  e^{-L_{\rm entry}/L_{D_2}} - e^{-(L_{\rm exit})/L_{D_2}} .
\ee
It depends on the distance between the LHC IP and the point at which the $D_2$ particle enters (exits) the detector $L_{\rm entry}$ $(L_{\rm exit})$, and the decay length of $D_2$ in the laboratory (LAB) frame, given by $L_{D_2}= c\,\tau_{D_2}\gamma_{D_2}\beta_{D_2}$. Here $\tau_{D_2}$ is the decay time of $D_2,$ while $\beta_{D_2}$ 
and $\gamma_{D_2}$ are respectively its speed in units of speed of light ($c$), and its Lorentz factor, in the LAB frame.
Clearly, $f_{\rm dec}$ vanishes for those trajectories of $D_2$ which do not intersect the detector. In the region of interest, the main contribution to the $D_2$ decay width comes from the decays $D_2 \to D_1 f \bar{f}$ into a pair of SM fermions. While the contribution into leptons is straightforward to compute, for quarks we need to worry about the validity of perturbative QCD. We follow the strategy outlined in\,\cite{Bertuzzo:2020rzo} and use a perturbative computation when $m_2 - m_1 > 2$ GeV, while in the opposite case we use chiral perturbation theory and vector mesons matrix elements to capture the hadronic decays. For more details, see\,\cite{Bertuzzo:2020rzo}.
Eq.\,\eqref{eq:Nsignal} is computed averaging ($\langle \cdot \rangle$) over all the possible kinematical configurations of the $D_2$ particles.
For this purpose, we produced a sample of $D_1 D_2$ events with \verb!MadGraph5_aMC@NLO!, that we then use to statistically evaluate eq.\,\eqref{eq:Nsignal}.
Finally, in eq.\,\eqref{eq:Nsignal}, the quantity $\epsilon_{\rm det}$ takes into account the efficiency for the reconstruction of the events (that we simply take as 100\%), and selection cuts. For the latter we impose a requirement on the energy of the visible final states employing the following strategy.
Using \verb!MadGraph5_aMC@NLO! we generate a sample of decays of $D_2\rightarrow D_1 l\bar{l} $, where $l$ is a lepton. We then perform the appropriate Lorentz boost to go from the $D_2$ frame to the LAB frame, we evaluate the sum of energies of the two leptons $E_{\rm vis}$, the \emph{visible} particles, and we impose $E_{\rm vis}>E_{\rm cut}.$
The fraction of the events satisfying this requirement leads to the quantity $\epsilon_{\rm det}$ in eq.\,\eqref{eq:Nsignal}. This procedure can be used to evaluate $\epsilon_{\rm det}$ for each $D_1 D_2$ event produced in our simulation.  
For FASER and FASER 2, we take $E_{\rm cut}=100$ GeV, mimicking the cut considered in\,\cite{Feng:2017uoz}, for AL3X we adopt $E_{\rm cut}=3$ GeV, as in\,\cite{Gligorov:2018vkc}, while for FACET $E_{\rm cut}=10$ GeV\,\cite{Cerci:2021nlb}.
For all the other experiments, we take $E_{\rm cut}=1$ GeV, which is in the range of energy thresholds considered in\,\cite{Curtin:2018mvb}. 
In the right panel of Fig.\,\ref{fig:SIDM_plots} we show how the sensitivity changes doubling these reference values of $E_{\rm cut}$, for the representative cases of FASER 2 and MATHUSLA, respectively forward and off-axis detectors.
We compute projected 95\% C.L. limits imposing that the number of signal events is larger than $N=3$. With this procedure we are assuming that backgrounds in the different experiments can be reduced at a negligible level, see e.g. the discussions in\,\cite{Feng:2017uoz,FASER:2019aik,Curtin:2018mvb}. Finally, we compute the DM relic abundance using the public tool \verb!micrOMEGAs5.2!\,\cite{Belanger:2020gnr}.

\subsection{Experimental geometries}
\label{sec:experiments}
We shall now describe the experimental facilities that we consider in our analysis. The geometry of the {\bf MATHUSLA} experiments is detailed in~\cite{MATHUSLA:2020uve}. The detector is delimited by:
\be\label{eq:Mathusla_geom}
68\,{\rm m} \leq z \leq 168\, {\rm m}\ , ~~~ 60\,{\rm m} \leq x \leq 80\, {\rm m}\ ,~~~ -50\,{\rm m} \leq y \leq 50\, {\rm m}\ , 
\ee
where the coordinate system is centered at the LHC interaction point, the $z$ axis is along the beam direction, and $x$ denotes the vertical to to the surface. 
For a given trajectory of the $\chi_2$ particle from our simulation, we evaluate whether the trajectory intersects the detector, and computes the quantities $L_{\rm entry}$ and $L_{\rm exit}.$ 
The total expected integrated luminosity is $3\,\rm{ab}^{-1}$.

We then consider the forward {\bf FASER} detector and its proposed upgrade {\bf FASER 2}\,\cite{Feng:2017uoz,FASER:2019aik}.
The FASER detector will be a cylinder with a radius of $R=10$ cm, a length of 1.5 m, 
located 480 m from the ATLAS IP. The relevant integrated luminosity will be $150\,\rm{fb}^{-1}.$
For FASER 2 we take $R=1$ m, a length of 10 m, a distance of 650 m from the IP, and an integrated luminosity of $3\,\rm{ab}^{-1},$\,\cite{FASER2private,FASER2talk}.

Concerning {\bf ANUBIS}\,\cite{Bauer:2019vqk}, a cylindrical detector displaced from the interaction point is proposed. For simplicity, we approximate the circular base of the cylinder with a square of equal area. We take an integrated luminosity of $3\,\rm{ab}^{-1}$.
Another implementation of the ANUBIS geometry is provided in\,\cite{Dreiner:2020qbi}, see also\,\cite{Hirsch:2020klk}. We checked that using that parametrization leads to sensitivity curves consistent with our results. 

The proposed {\bf CODEX-b} would be installed next to the LHCb IP\,\cite{Gligorov:2017nwh,Aielli:2019ivi}. Its geometry is defined by 
\be\label{eq:Codexb_geom}
5\,{\rm m} \leq z \leq 15\, {\rm m}\ , ~~~ 26\,{\rm m} \leq x \leq 36\, {\rm m}\ ,~~~ -7\,{\rm m} \leq y \leq 3\, {\rm m}\ , 
\ee
and the total luminosity would be $300\,\rm{fb}^{-1}.$

We then consider the {\bf AL3X} detector, proposed to be placed close to the ALICE experiment at the LHC\,\cite{Gligorov:2018vkc}. It is a 12 m long cylinder centered around the beam axis about 4.25 m from the IP with inner and outer radii respectively of 0.85 m and 5 m. This forward detector will cover the pseudorapidity range $0.9\lesssim \eta \lesssim 3.7$. 
We take an integrated luminosity of $250\,\rm{fb}^{-1},$ which correspond to the optimistic value proposed in\,\cite{Gligorov:2018vkc}.

The {\bf MAPP-1} and its upgrade {\bf MAPP-2} detectors will be placed in the UGCI gallery adjacent to the MoEDAL region\,\cite{Staelens:2019gzt,Pinfold:2019zwp}. The projected integrated luminosity for these facilities is respectively of $30\,\rm{fb}^{-1}$ and $300\,\rm{fb}^{-1}.$ Concerning their geometry, we adopt the same setup of\,\cite{Dreiner:2020qbi}.

Very recently, a forward detector, {\bf FACET}, to be placed around the CMS interaction point, has been proposed\,\cite{Cerci:2021nlb}. The detector volume is 18 m long, covering $101\,{\rm m}<z<119\,{\rm m},$ and it is sensitive to particles with polar angles $1\,{\rm mrad}<\theta<4\,{\rm mrad}.$ The expected total luminosity received is $3\,\rm{ab}^{-1}.$

\section{Results}
\label{sec:results}
\begin{figure}[tb]
\includegraphics[width=0.48\textwidth]{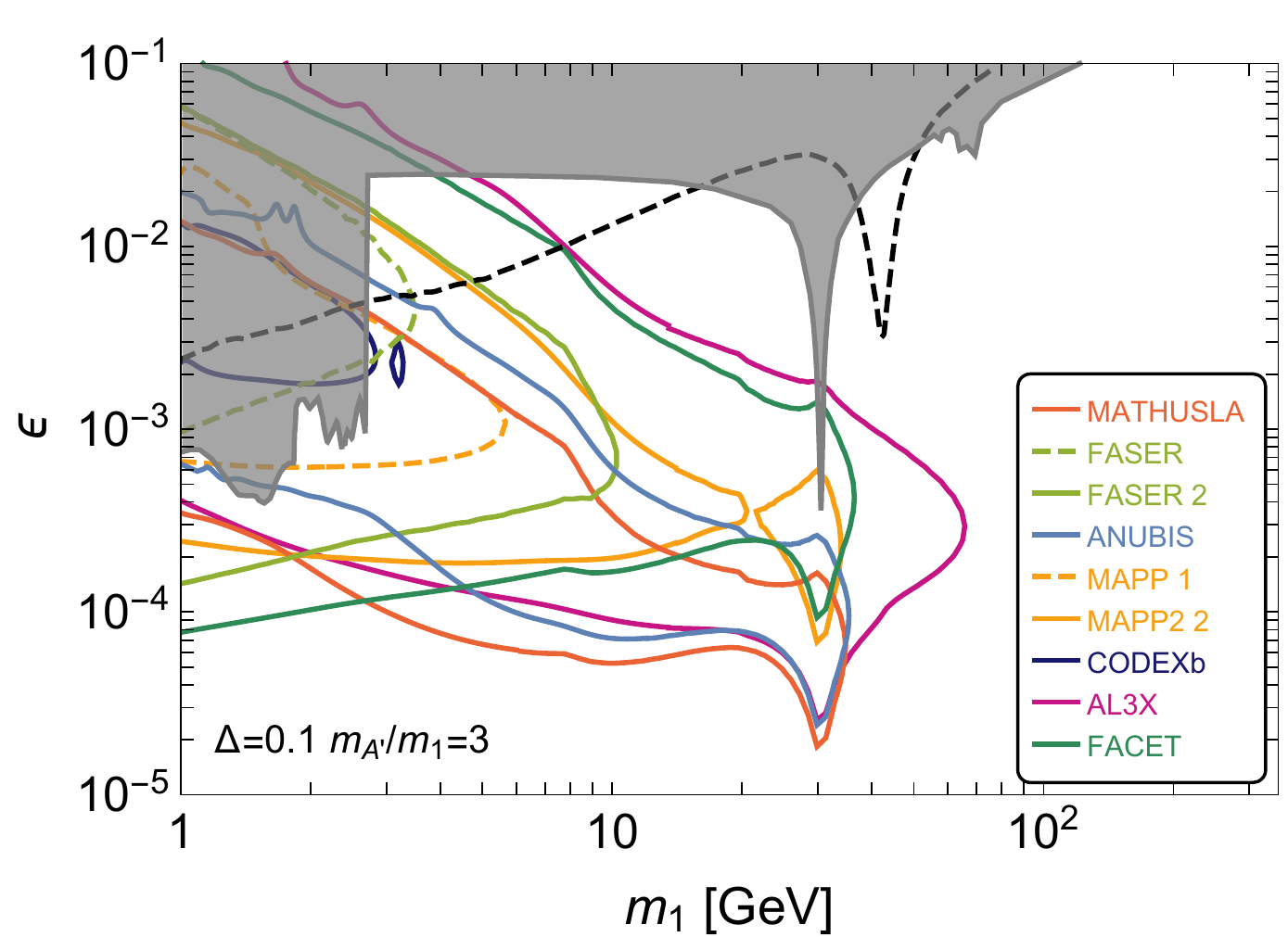}
\quad
\quad
\includegraphics[width=0.48\textwidth]{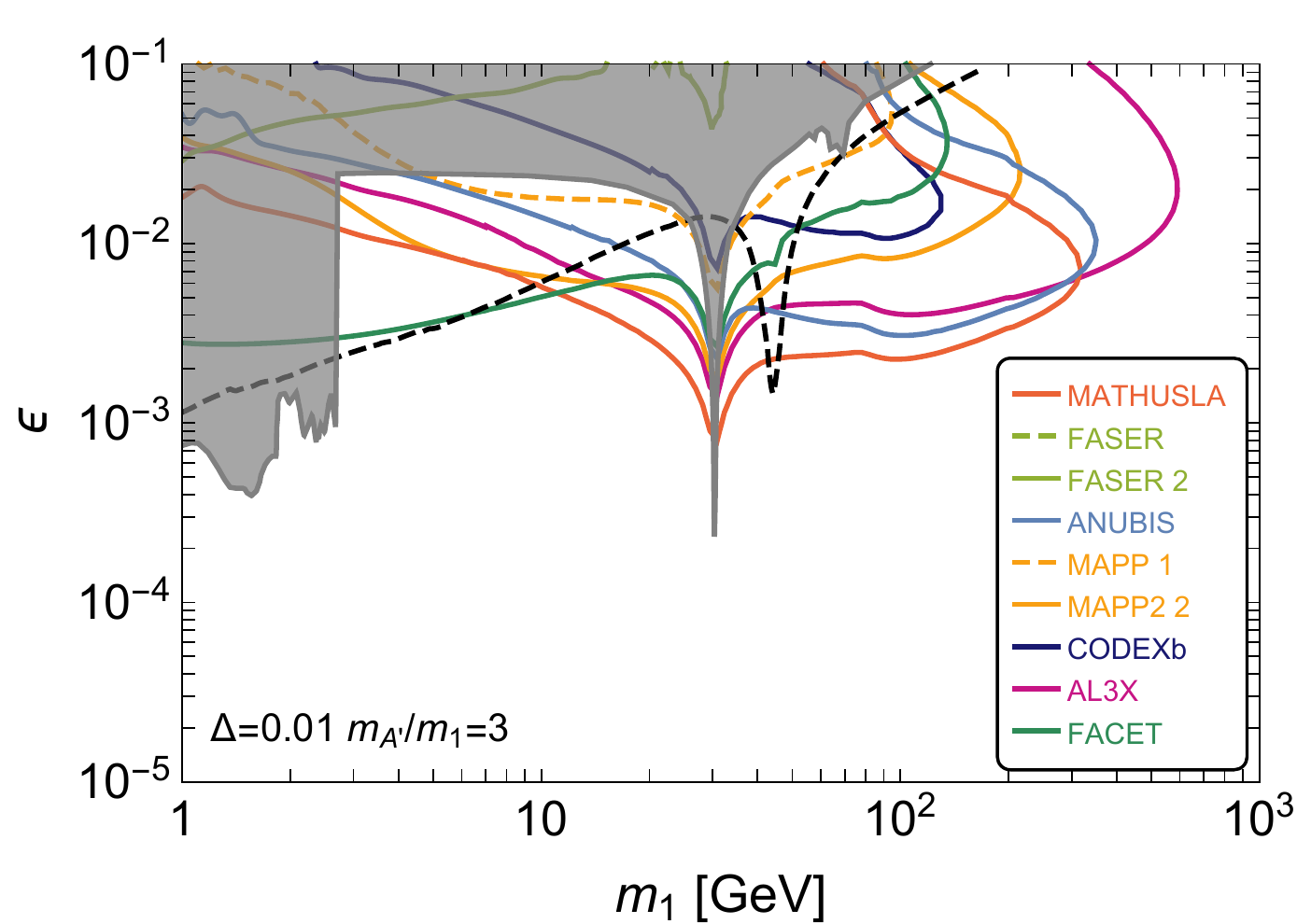}\\
\includegraphics[width=0.48\textwidth]{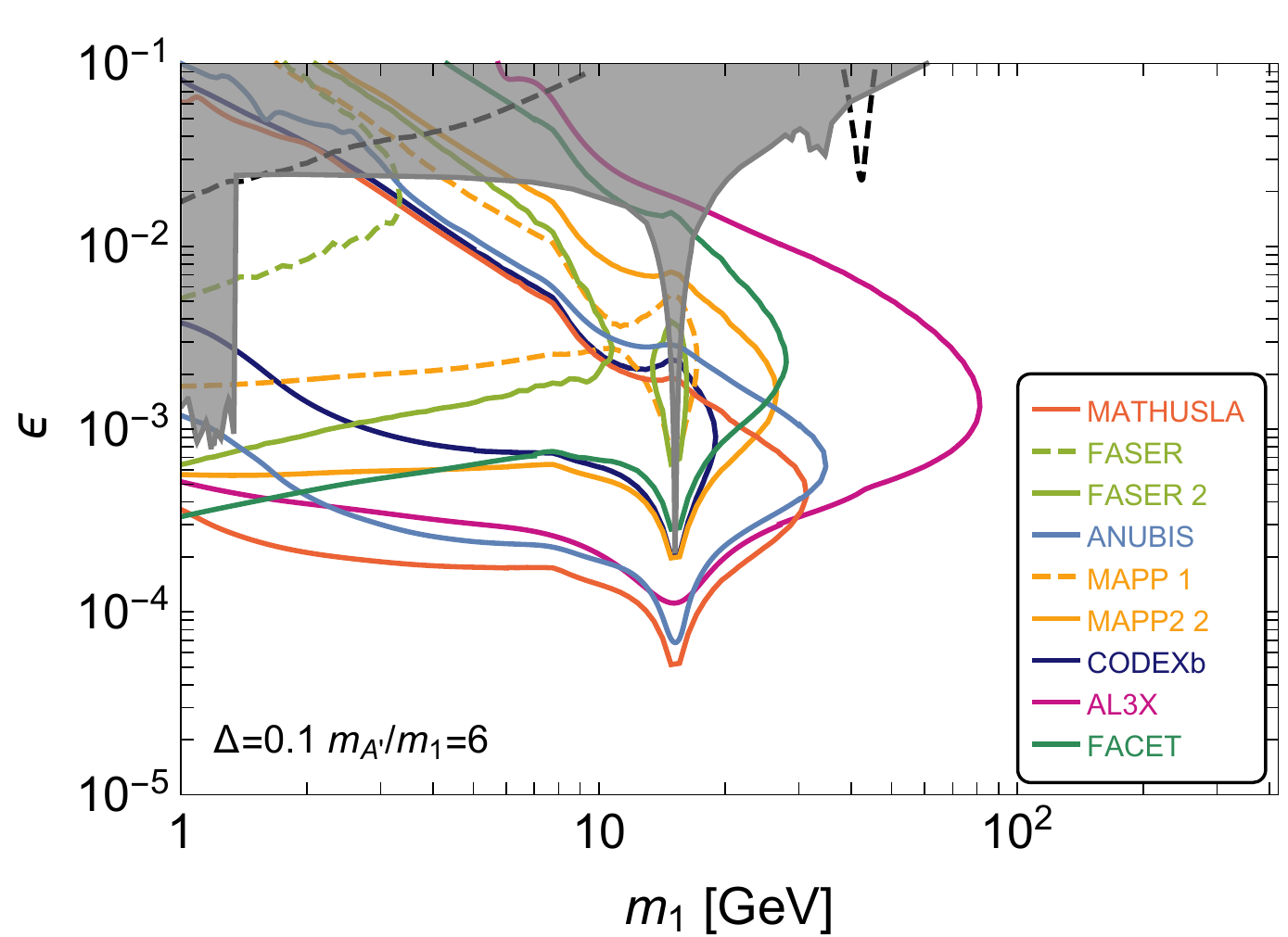}
\quad
\quad
\includegraphics[width=0.48\textwidth]{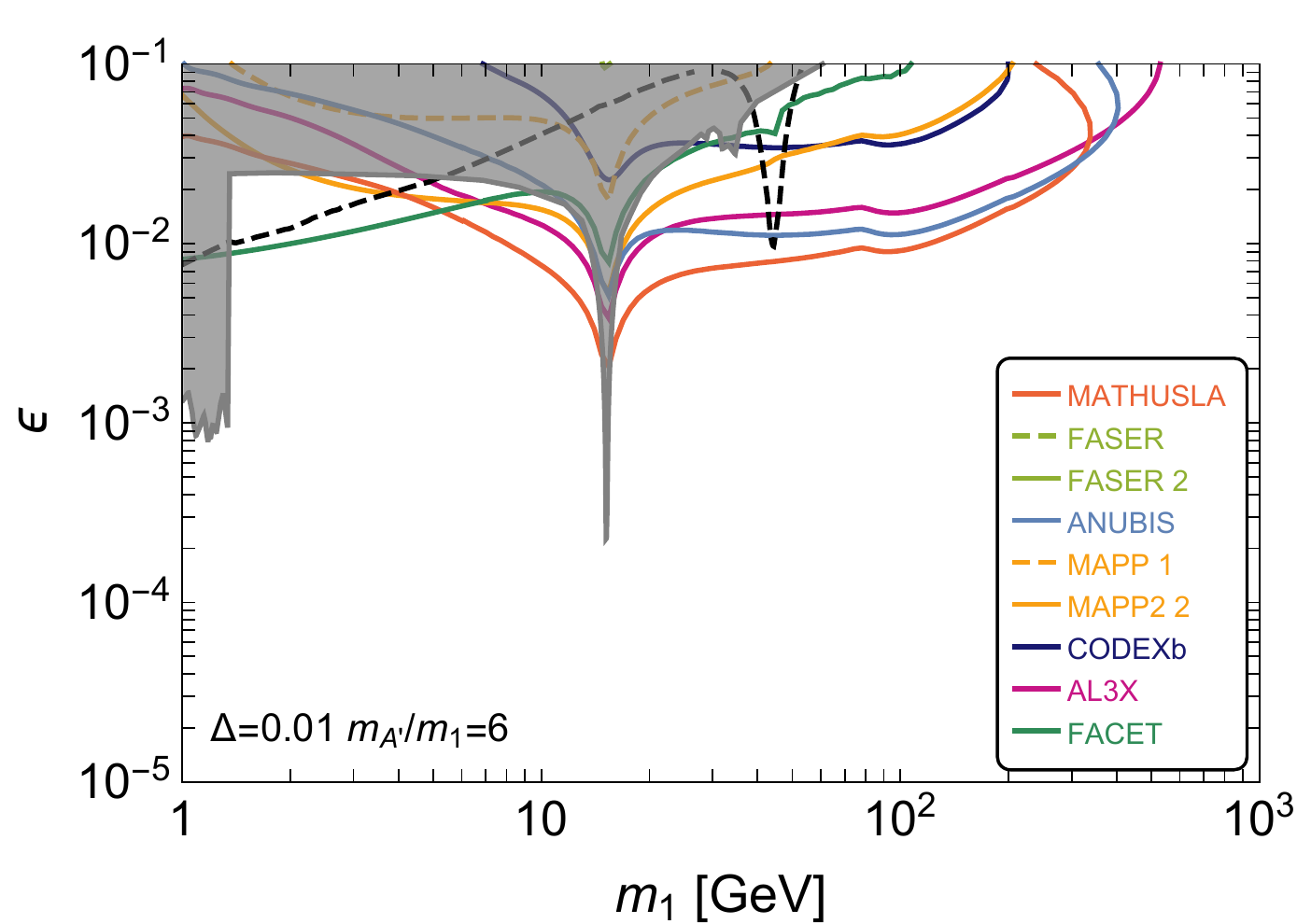}
\caption{Projected sensitivities of proposed future experiments (coloured lines) and existing constraints (grey regions) on the model of fermionic iDM described in Sec.~\ref{sec:FIDM_summary} in  the $m_1-\epsilon$ plane for $\Delta = \left\{0.1, 0.01 \right\}$ and $m_{A'}/m_1 = \left\{3, 6\right\}$. The dashed black contour depicts where the abundance of the $\chi_1$ matches the observed dark matter energy density. 
}
    \label{fig:FIDM_plots}
\end{figure}
\begin{figure}[tb]
    \centering
   \includegraphics[scale=0.48]{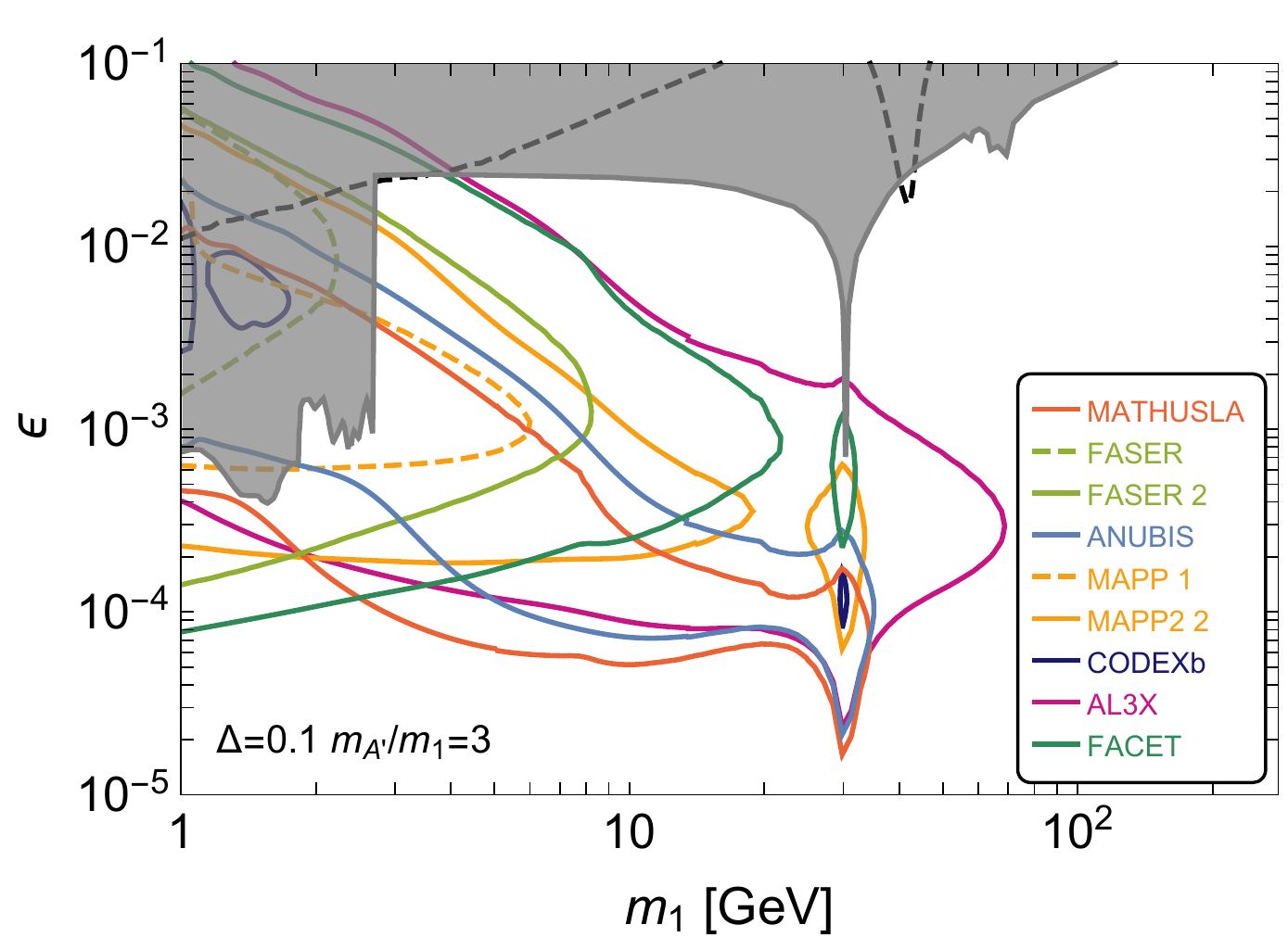}
\quad
\quad
\includegraphics[scale=0.48]{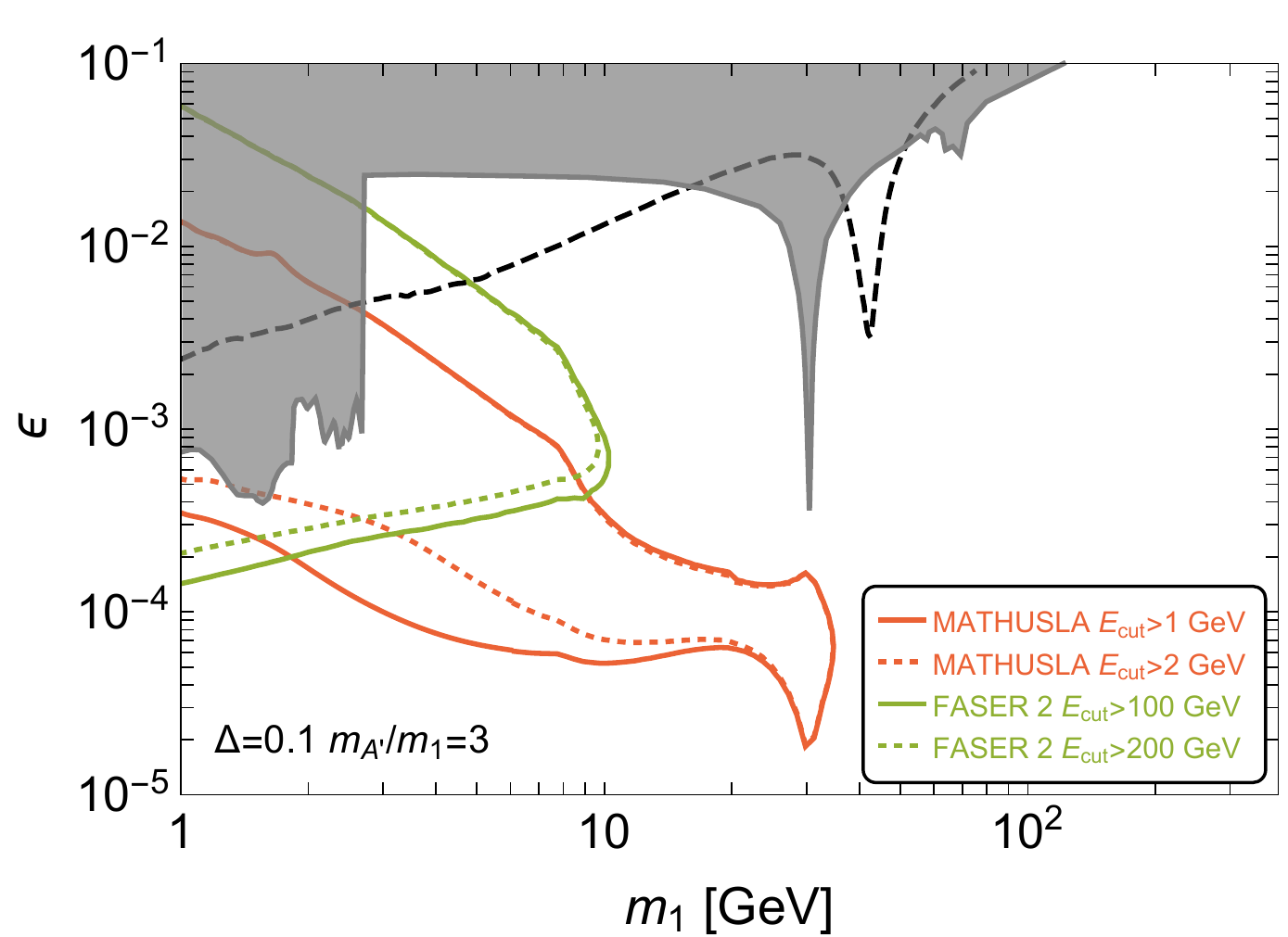}
\caption{Left panel: The same as Fig.~\ref{fig:FIDM_plots} but for the case of the scalar iDM model described in Sec.~\ref{sec:SIDM_summary}. For illustration, we only show the case of the first benchmark considered ($\Delta = 0.1, m_{A^\prime}/m_1 = 3$) as the results between the scalar and fermionic cases do not differ drastically for most of the experiments. Right panel: Projected sensitivities of MATHUSLA and FASER 2 for the fermionic iDM model and for two choices of the energy threshold $E_{\rm cut}.$
}
    \label{fig:SIDM_plots}
\end{figure}

In this section we present the results of the forecasted sensitivities using the methods and experiments detailed in Sec.\,\ref{sec:setup}. 
These are shown in the $m_1 - \epsilon$ plane for DM masses $m_1 \geq 1$ GeV. This region is interesting because ample areas are unconstrained by current experiments and, as we are going to see, the LHC experiments listed in sec.\,\ref{sec:experiments} can probe large portions of it. In the same region we can also have thermal DM production. However, we do not restrict only to the parameters that allow to obtain the measured DM thermal abundance, since we can imagine a non-thermal DM production or a modified cosmological history that could dramatically change the picture. In all our plots we fix $\alpha_D = g_d^2/(4\pi) = 0.1$.
To illustrate the complementarity of the different future LHC experiments, we focus on four representative benchmark scenarios, considering two different values for the mass splitting ($\Delta = 0.1$ and $0.01$) and for the ratio between the dark photon and the DM mass ($m_{A'}/m_1 = 3$ and $6$). 
The main results are presented in Fig.\,\ref{fig:FIDM_plots} and Fig.\,\ref{fig:SIDM_plots}.
The shaded grey region depicts the current experimental constraints on the invisible dark photon already mentioned in Sec.\,\ref{sec:model}, coming from BaBar\,\cite{BaBar:2017tiz} and LEP\,\cite{Curtin:2014cca}. Since in the region $m_{A'} \sim m_Z$ the $Z$ boson coupling to the dark states is not suppressed, we also include the bound coming from the $Z$ invisible decay\,\cite{Zyla:2020zbs}. 
The colored contours, on the other hand, show the projected future sensitivities of the experiments listed in Sec.~\ref{sec:experiments}. Finally, along the dashed black line the $D_1$ abundance matches the observed one. 
Fig.~\ref{fig:FIDM_plots} and Fig.~\ref{fig:SIDM_plots} correspond to the fermionic iDM (see Sec.\,\ref{sec:FIDM_summary}) and scalar iDM (see Sec.\,\ref{sec:SIDM_summary}) models, respectively.

The forecasted sensitivities are very similar between the fermionic and scalar cases for most of the experiments. 
This can be understood in the following way: on the one hand, the lifetimes of $\chi_2$ and $\phi_2$ are very similar, and 
the same is true for production of fermionic and scalar iDM pairs from on-shell dark photon decays.
On the other hand, events with very off-shell dark photons 
give rise to different
energy distributions 
depending 
on the dark states spin, and this explain the
differences observed among the two scenarios. For this reason, in Fig.\,\ref{fig:SIDM_plots} we only show the benchmark $\Delta = 0.1$ and $m_{A'}/m_1 = 3$. 
Concerning the predictions for the DM relic abundance, the difference between the fermionic and scalar cases is instead more pronounced, due to the different velocity dependence of the non-relativistic annihilation cross-section.

As shown in Fig.~\ref{fig:FIDM_plots} and Fig.~\ref{fig:SIDM_plots}, we find that most of the experiments will be able to probe regions of parameter space not excluded by current data and in which $D_1$ can be $100\%$ of the DM. More specifically we see that, for $\Delta = 0.1$, the thermal line will be completely probed for $m_1 \lesssim 10$ GeV, while for $\Delta = 0.01$ the thermal line will be approximately covered in all the parameter space considered.

Comparing the $\Delta = 0.1$ and $0.01$ cases, we see that for the latter choice the sensitivity of LHC experiments moves to larger values of $m_1$ and $\epsilon$. This can be qualitatively understood looking at the decay width of $D_2 \to D_1 f \bar{f}$. In the limit of massless fermions and for small $\Delta$ we have $\Gamma \propto \epsilon^2 \Delta^5 m_1^5/m_{A'}^4$\,\cite{Berlin:2018jbm}.
A decrease of $\Delta$ can be compensated with an increase of $\epsilon$ and $m_1$ to obtain the same decay width (and hence lifetime), reproducing the behaviour observed.
Instead, moving from $\Delta = 0.1$ and $0.01$ has only a modest impact on the predictions for the DM relic abundance.
The situation is different for what concerns the role of the ratio between the dark photon and the DM mass. Going from $m_{A'}/m_1 = 3$ to $m_{A'}/m_1 = 6$ reduces the impact of the dark photon resonance for the DM annihilations in the early Universe. This implies that larger values of $\epsilon$ should be considered to obtain the same DM relic abundance.
Let us also 
stress once more
that the predictions for the DM abundance could be completely different in presence of non-thermal production mechanisms or scenarios of non-standard cosmology.

Concerning the different experiments, as evident from the discussion in Sec.\,\ref{sec:setup}, the sensitivity depends on a combination of several ingredients: the proximity of the detectors to the IP, their off-axis vs on-axis geometry, their size, luminosity received, and other factors such as the selection cuts performed in the analysis. As shown in Fig.~\ref{fig:FIDM_plots} and Fig.~\ref{fig:SIDM_plots}, the different facilities considered here are complementary, but at the same time they cover overlapping regions of the parameter space.

{The sensitivity reach of the AL3X experiment is particularly remarkable, although the feasibility of this proposal strongly depends on the overlap with the ALICE physics program~\cite{Gligorov:2018vkc}.
The ANUBIS experiment allows to reach sensitivities comparable to the ones for MATHUSLA despite the smaller detector size, a factor which is compensated by its proximity to the IP.
From Fig.~\ref{fig:FIDM_plots} and Fig.~\ref{fig:SIDM_plots} one can notice that MATHUSLA typically probes slightly smaller values of $\epsilon,$ but ANUBIS allows to test complementary parts of the parameter space (see e.g. the region around $m_1\sim 20$ GeV in Fig.~\ref{fig:SIDM_plots}).
The recently proposed forward detector FACET also offers good sensitivities. The main advantage with respect to the FASER 2 is its proximity to the IP (101 m for FACET and 650 m for FASER 2) and the larger solid angle coverage. These features allow an increased sensitivity with respect to FASER 2, at least as long as the background can be reduced to negligible levels as assumed in the original proposal~\cite{Cerci:2021nlb}.
The projected reach of MAPP 2 is complementary to the ones previously mentioned, despite the smaller integrated luminosity that this detector will receive (0.3 ab$^{-1}$ instead of 3 ab$^{-1}$). At masses $m_1\gtrsim 10$ GeV, the sensitivity curves are qualitatively similar to the ones for MATHUSLA, but shifted to larger values of $\epsilon.$ 
Overall, these facilities are complementary to CODEXb, FASER (FASER 2) and MATHUSLA, previously considered in~\cite{Berlin:2018jbm}, surpassing their sensitivities in some regions of the parameter space.
At qualitative level, similar results have been found for different models, see e.g.~\cite{Cerci:2021nlb,Dreiner:2020qbi,Hirsch:2020klk,Deppisch:2019kvs}. However, the comparison of one particular experiment over the others  strongly depends on the physical scenario under consideration, motivating therefore our dedicated study for iDM.
Concerning CODEXb, FASER 2 and MATHUSLA, we have found results similar to~\cite{Berlin:2018jbm}, once the same setup is adopted. For this study, we have used the most updated geometry proposals for FASER 2 and MATHUSLA. These lead to slightly decreased performances for the former detector, and increased sensitivities for the latter (especially around $m_1\simeq$ 20 GeV for the scenarios with $\Delta=0.1$ and $m_{A^{\prime}}/m_1=3$).

}

Finally, we shall mention that searches of iDM complementary to those analyzed here can be performed at ATLAS, CMS and LHCb~ looking for displaced muons and time-delayed tracks~\cite{Berlin:2018jbm}, and at Belle II with displaced vertices and missing momentum signatures~\cite{Duerr:2019dmv,Duerr:2020muu}.

\section{Conclusions}
\label{sec:conclusions}

We have considered a model of fermionic (and scalar) iDM coupled to the SM via a dark photon mediator. In this simple scenario, even for relatively light DM,  it is possible to obtain the observed cosmological DM abundance via the standard freeze-out mechanism and, at the same time, fulfill the constraints from direct and indirect DM searches and CMB observations. 
The heaviest iDM state can be long-lived and it can be searched for at proposed LHC experiments. In this paper, we have updated and extended the analysis in\,\cite{Berlin:2018jbm}, considering the future detectors FASER, MATHUSLA, CODEX-b, AL3X, MAPP, ANUBIS and FACET.
The main results are shown in Fig.\,\ref{fig:FIDM_plots} and Fig.\,\ref{fig:SIDM_plots}.
We find that the experimental facilities discussed here offer promising prospects for detection. They can cover complementary regions of the parameter space, significantly extending the reach of current and past experiments. In the case of the mass ratio $m_{A'}/m_1=6$, and for the range of $\Delta$ considered $0.01<\Delta<0.1$, they will be able to 
completely probe the parameter space in which the lightest state constitutes $100\%$ of the DM via the standard thermal freeze-out scenario.

\medskip

\section*{Acknowledgements}
We thank M. Hirsch for useful discussions, as well as A. Berlin, D. Curtin, J. Feng, F. Kling, S. Trojanowski and Z. S. Wang for correspondence about their work on the experiments considered. 
EB acknowledges financial support from FAPESP under contract 019/04837-9. M.T. acknowledges support from the INFN grant ``LINDARK''. A.S. and M.T. acknowledge the research grant ``The Dark Universe: A Synergic Multimessenger Approach No. 2017X7X85'' funded by MIUR, and the project ``Theoretical Astroparticle Physics (TAsP)'' funded by the INFN.

\medskip
\small

\medskip
\small

\bibliographystyle{JHEP}
\bibliography{DM_bib}
\end{document}